\documentstyle[aps,epsfig,floats]{revtex}

\begin{document}
\draft
\def\dbltopfraction{1.0}

\wideabs{%must remove this before submitting to APS!!  
\title{Spin Injection in a Ballistic Two-Dimensional Electron Gas}
\author{H.X. Tang, F.G. Monzon, Ron Lifshitz, M.C. Cross, and
M.L. Roukes\cite{roukes}}
\address{Condensed Matter Physics 114-36, California Institute of
Technology, Pasadena, CA 91125, USA}
\date{\today} 
\maketitle

\begin{abstract}
We explore electrically injected, spin polarized transport in a
ballistic two-dimensional electron gas. We augment the
B\"{u}ttiker-Landauer picture with a simple, but realistic model for
spin-selective contacts to describe multimode reservoir-to-reservoir
transport of ballistic spin 1/2 particles. Clear and unambiguous
signatures of spin transport are established in this regime, for the
simplest measurement configuration that demonstrates them
directly. These new effects originate from spin precession of
ballistic carriers; they exhibit strong dependence upon device
geometry and vanish in the diffusive limit. Our results have important
implications for prospective ``spin transistor'' devices.
\end{abstract}
}%end wideabs

\begin{figure}[t]
\begin{center}
\epsfig{file=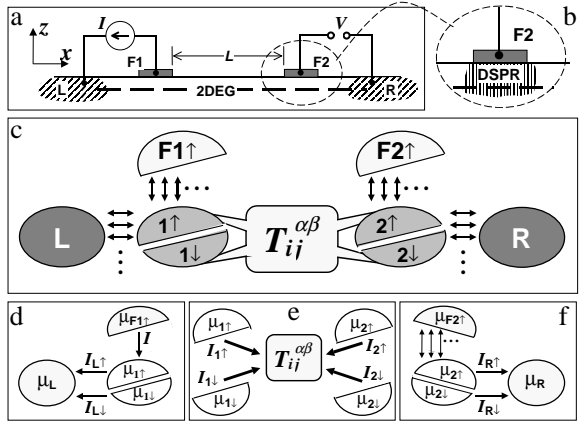,width=3.0in}
\end{center}
\caption{Model for ballistic spin injection in 2D. (a) Measurement
configuration: a current, $I$, is injected through the 2DEG via a
ferromagnetic contact {\bf F1} and an ohmic contact {\bf L}. The
spin transresistance, $R_S = V/I$, arises from spin polarized carriers
traversing a distance $L$ from the path of the net current, which induce a
nonlocal voltage, $V$, between a second, similar, pair of contacts,
{\bf F2} and {\bf R}. (b) The conductor beneath the ferromagnetic
contacts (DSPR) is assumed to be a {\it Disordered,} but {\it Spin
Preserving Region.} (c) The full 8-reservoir model; complete ellipses
represent spin-relaxing reservoirs, half ellipses represent
spin-resolved reservoirs. {\bf F1} and {\bf L} are current contacts,
{\bf F2} and {\bf R} are voltage probes. $T_{ij}^{\alpha\beta}$
denotes the 2DEG device channel in which spin precession occurs. Other
multimode leads are denoted by three arrows and ellipsis. Panels (d),
(e), (f) illustrate decomposition of the 8-reservoir model. (Panel (e)
depicts the reduced 4-reservoir problem.)}
\label{fig1}
\end{figure}

The concept of a spin transistor, first proposed almost ten years
ago~\cite{1}, has attracted widespread interest~\cite{2} but its
experimental realization remains elusive~\cite{3}. It is based upon
electrical injection of spin polarized carriers from a ferromagnetic
conductor into an electron gas within a semiconductor. Electrons
propagating in the interfacial electric field confining them to the
device channel experience an effective magnetic field that induces
spin precession; this is called the Rashba effect~\cite{4}. The rate
of spin precession should be tunable through an external gate voltage,
which will add to the confinement potential~\cite{5,6} With
ferromagnetic source and drain contacts acting as spin polarizer and
analyzer, an electron device analogous to an electro-optic
modulator~\cite{7} is envisaged.

Experiments to date demonstrate that the spin transistor geometry
[Fig.~\ref{fig1}(a)], in general, leads to strong Hall phenomena that
are unrelated to true spin transport~\cite{3}. They arise due to the
requisite proximity of miniature magnets and the low density (and,
hence, high Hall coefficient) electron gas. Since these Hall phenomena
depend directly upon the magnetization state of these magnetic
contacts, they often closely mimic the signals expected from spin
transport experiments---especially those in which the relative
magnetic orientation of the ``spin polarizer'' and ``analyzer''
contacts is varied. However, in early experiments on (diffusive) spin
injection in metals~\cite{8}, spin precession phenomena provided an
alternate and crucial experimental proof of spin transport. In this
Letter, we establish for the first time the analogous, and
unambiguous, experimental signatures to be expected from spin
injection and precession phenomena in a ballistic two-dimensional
electron gas (2DEG). The observation of these precessional effects
will constitute a definitive experimental demonstration of electrical
spin injection in semiconductor systems.

The spin transresistance, $R_S$ [Fig.~\ref{fig1}(a)] provides the most
direct demonstration of spin transport~\cite{8}. This nonlocal
transport coefficient is free of obfuscating background signals
unrelated to spin injection. In the {\it diffusive\/} limit, if
current contact {\bf F1} is replaced with one that is unpolarized, no
voltage will appear between the analyzer contact {\bf F2} and a
suitably defined ground reference {\bf R}.  In this case these voltage
contacts, being well outside the net current path, remain at
equipotential.  With a polarized current contact {\bf F1}, injected
magnetization can lead to steady-state {\it spin accumulation\/} that
persists over the entire length of the channel if $\delta_S \gtrsim L$. This
induces disequilibrium between the local spin-resolved electrochemical
potentials of {\bf F2} and {\bf R}, and yields a finite $R_S$.  Here
$\delta_S = \sqrt{\ell_0 \ell_S/2}$ is the spin diffusion length,
$\ell_S = v_F \tau_S$ and $\ell_0=v_F \tau_0$ are the spin and
momentum mean free paths, $\tau_S$ and $\tau_0$ are the effective spin
and momentum relaxation times, and $v_F$ is the Fermi velocity.

In the {\it ballistic\/} regime, however, it is not appropriate to
speak of spin accumulation since a local chemical potential cannot be
meaningfully defined within the channel. Accordingly, our description
of the ballistic spin transresistance is based upon B\"{u}ttiker's
picture for mesoscopic transport within a multiprobe
conductor~\cite{9}. Here we augment this with a new model describing
spin-selective contacts. Our procedure is as follows: (a) We first
develop a simple description of spin-selective contacts, based upon
careful consideration of the ferromagnetic/semiconductor (F/S)
contacts in (our) real devices~\cite{3}. (b) We construct an
8-reservoir model, after B\"{u}ttiker, to describe the spin injection
experiment. (c) Boundary conditions are used to constrain the
spin-resolved currents and chemical potentials. These lead to a
simpler 4-reservoir problem for the spin transresistance, $R_S$, in
terms of reservoir-to-reservoir, {\it spin-resolved\/} transmission
probabilities, $T_{ij}^{\alpha\beta}$, of the 2DEG forming the device
conduction channel. Here the indices $i$, $j$ specify the reservoirs
themselves, and $\alpha$, $\beta$ their constituent spin bands. (d)
The requisite $T_{ij}^{\alpha\beta}$ are then calculated
semiclassically, using a modified Monte Carlo numerical technique
(described below).  We follow the electrons' ballistic trajectories
{\it and\/} the phase of their spin wavefunctions as they pass through
the device, while ignoring the phase of their spatial
wavefunctions. For unpolarized ballistic systems, this semiclassical
approach has proven remarkably consistent with experimental data at
$T\sim 4$K, where the electron phase coherence length is smaller than
typical dimensions of nanoscale devices~\cite{10}.

Figure~\ref{fig1}(b) depicts our model for the spin-selective contacts
which comprises two elements: {\bf F2}, a fully spin-polarized
reservoir which is in perfect contact with DSPR, a second {\it
Disordered\/} ({\it i.e.}\ momentum-randomizing) but {\it
Spin-Preserving Region\/} that consists of separate spin-up and
spin-down bands. The separate spin-resolved reservoirs comprising the
DSPR's (1$\uparrow$, 1$\downarrow$, 2$\uparrow$, 2$\downarrow$) model
low mobility regions always present beneath unalloyed ferromagnetic
metal contacts in typical InAs devices~\cite{3}. Disorder within them
yields significant momentum randomization and, hence, a short
$\ell_0$. However, in contrast to the usual picture describing
unpolarized reservoirs~\cite{9}, we assume these special contacts are
small compared to $\delta_S$, thus any spin disequilibrium within
them is preserved. This, in fact, is consistent with the more
restrictive constraint, $\delta_S\gtrsim L$, which is generic and
fundamental {\it to any spin injection experiment.} If significant
spin relaxation occurs anywhere in the device, including the vicinity
of the ferromagnetic contacts, spin-selective transport is
suppressed~\cite{13}. In this Letter, for sake of clarity, we
consider the most ideal situation, initially assuming that {\bf F1}
and {\bf F2} are fully polarized at the Fermi surface (half-metals).
This approximation serves to illustrate the most important aspects of
the underlying physics. Of course, many complexities in real devices
may diminish spin transport effects~\cite{AA}. Here our aim is to
establish what may be expected in ballistic systems under {\it
optimal\/} conditions.

Measurement of $R_S$ involves four terminals [Fig.~\ref{fig1}(a)], two
that are spin-selective, {\bf F1} and {\bf F2}, and two that are
conventional, {\it i.e.}\ momentum- {\it and\/} spin-relaxing, {\bf
L} and {\bf R}. As depicted in Figures~\ref{fig1}(d, e, f), the full
problem separates into three sub-components. Fig.~\ref{fig1}(d)
represents the spin-up and spin-down currents ($I_{L\uparrow}$,
$I_{L\downarrow}$) that flow between {\bf F1}, 1$\uparrow$,
1$\downarrow$, and {\bf L}. A Sharvin resistance~\cite{15},
$R_{sh}=(h/2e^2)(k_F w)/\pi = (h/2e^2)N_{ch}$, arises between
1$\uparrow$, 1$\downarrow$ and the multichannel conductors connecting
them to {\bf L}. Under conditions of current flow this yields the
spin-resolved electrochemical potential differences
$\mu_{1\uparrow}-\mu_L = 2eR_{sh}I_{L\uparrow}$ and
$\mu_{1\downarrow}-\mu_L = 2eR_{sh}I_{L\downarrow}$. Here, the factors
of 2 arise because transport is spin resolved; $k_F$, $w$, and
$N_{ch}$ are the Fermi wave vector, channel width, and number of
occupied modes within the 2DEG device channel, respectively.
Similarly, at the rightmost side of Fig.~\ref{fig1}(f), current flow
between the reservoirs 2$\uparrow$, 2$\downarrow$ and {\bf R}
establishes the electrochemical potential differences
$\mu_{2\uparrow}-\mu_R = 2eR_{sh}I_{R\uparrow}$ and
$\mu_{2\downarrow}-\mu_R = 2eR_{sh}I_{R\downarrow}$. Also,
$\mu_{F2\uparrow}= \mu_{2\uparrow}$ since no current flows between
these reservoirs. Note that all $I$'s here represent {\it net\/}
currents (forward minus reverse contributions). In our model, the
following sum rules hold: $I=I_{L\uparrow} + I_{1\uparrow}$,
$I=I_{L\uparrow} + I_{L\downarrow}$, $0=I_{R\uparrow} +
I_{R\downarrow}$, and $I_{1\uparrow} + I_{1\downarrow} = I_{2\uparrow}
+ I_{2\downarrow} =0$.  As the reservoirs in Fig.~\ref{fig1}(f) are
voltage contacts, net current is conserved separately for each spin
band, $I_{R\uparrow} + I_{2\uparrow} = I_{R\downarrow} +
I_{2\downarrow} =0$. These expressions can be manipulated to yield
\begin{equation}\label{eq1}
\pmatrix{
\mu_{1\uparrow}\cr \mu_{1\downarrow}\cr \mu_{2\uparrow}\cr \mu_{2\downarrow}\cr
} =
\pmatrix{
\mu_L + 2eR_{sh}(I-I_{1\uparrow})\cr
\mu_L + 2eR_{sh}I_{1\uparrow}\cr
\mu_R - 2eR_{sh}I_{2\uparrow}\cr
\mu_R + 2eR_{sh}I_{2\uparrow}\cr
}.
\end{equation}

\begin{figure}[t]
\begin{center}
\epsfig{file=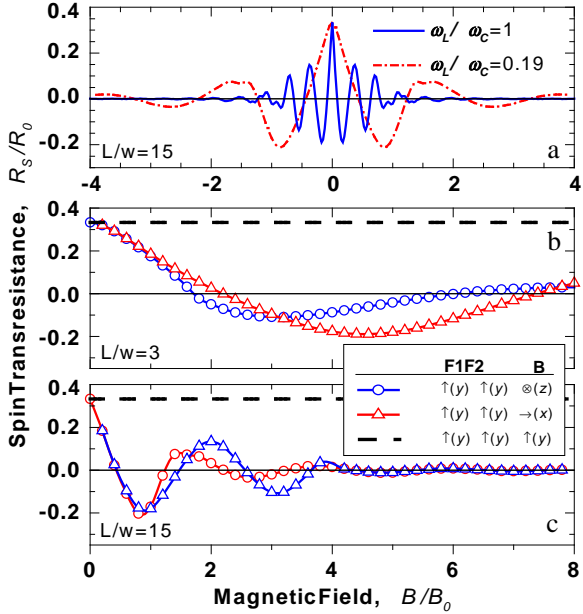,width=3.0in}
\end{center}
\caption{Ballistic spin transresistance in an external field
normalized to $B_0=p_F/ew$, at which the cyclotron radius equals the
channel width. (a) For a channel with $L/w=15$ in perpendicular field,
we plot two traces representing $\omega_L/\omega_c =1$ and 0.19,
appropriate for a typical metal and for InAs, respectively. (b,c) Spin
transresistance for three different configurations and two channel
lengths, $L/w=3$ and 15.  Here $\omega_L/\omega_c=0.19$ (InAs).}
\label{fig2-3}
\end{figure}

Given these relations, calculation of $R_S$ reduces to a 4-terminal
problem that solely involves the four spin-resolved reservoirs:
1$\uparrow$, 1$\downarrow$, 2$\uparrow$, and 2$\downarrow$ and the
2DEG device channel that connects them [Fig.~\ref{fig1}(e)]. Modifying
B\"{u}ttiker's formula to account for the spin-resolved channels, the
4-terminal linear response at zero temperature becomes
\begin{equation}\label{eq2}
I_{i\alpha} = {e\over h} \left[(N_{ch} - R_{ii}^{\alpha\alpha})\mu_{i\alpha} - 
T_{ij}^{\alpha\beta}\mu_{j\beta} \right]
\equiv  {e\over h} U_{ij}^{\alpha\beta}\mu_{j\beta}.
\end{equation}

Transport within the ballistic multimode 2DEG conductor is fully
represented by the transmission and reflection coefficients,
$T_{ij}^{\alpha\beta}$ and $R_{ii}^{\alpha\alpha}$. These describe
carriers incident from the lead $i$ with spin polarization $\alpha$,
that are transmitted into lead $j$ with final spin state $\beta$ ; and
carriers incident from $i\alpha$ that are reflected back into same
lead and spin channel, respectively. The coefficients $U_{ij}$ in
Eq.~\ref{eq2} satisfy the sum rule $\sum_{i\alpha}
U_{ij}^{\alpha\beta}= \sum_{j\beta} U_{ij}^{\alpha\beta} =0$ ensuring
the current sum rules of Eq.~\ref{eq1}, and that all currents vanish
when the $\mu_i$ are equal.

Simplification of Eq.~\ref{eq1} and \ref{eq2} yields
\begin{equation}\label{eq3}
\pmatrix{
I_{1\uparrow}\cr I_{1\downarrow}\cr 
I_{2\uparrow}\cr I_{2\downarrow}\cr
} =  \bbox{S}
\pmatrix{
\tilde{\mu}_L + I\cr
\tilde{\mu}_L\cr
\tilde{\mu}_R\cr
\tilde{\mu}_R\cr
}.
\end{equation}
where $\tilde{\mu}_{L,R}=\mu_{L,R}/2eR_{sh}$ and $\bbox{S}\equiv
(1+\bbox{U})^{-1}\bbox{U}$. The elements of $\bbox{S}$ satisfy the
same sum rules that constrain $\bbox{U}$ (for identical reasons). For
parallel alignment of polarizer and analyzer, {\bf F1} and {\bf F2},
which we denote by the superscript ($\uparrow\uparrow$), these steps
yield
\begin{equation}\label{eq4}
R_S^{(\uparrow\uparrow)} = -2 {{S_{31}S_{42} - S_{32}S_{41}} \over
{S_{31} + S_{32} + S_{41} + S_{42}}}
R_{sh}.
\end{equation}
For antiparallel alignment, only the sign changes:
$R_S^{(\uparrow\downarrow)} = -R_S^{(\uparrow\uparrow)}$.

We obtain the requisite elements of $\bbox{S}$ numerically, extending
the semiclassical billiard model~\cite{10} to allow tracking of an
electron's spin wavefunction along ballistic trajectories linking the
spin-resolved reservoirs
(1$\uparrow$,1$\downarrow$,2$\uparrow$,2$\downarrow$) at either end of
the 2DEG device channel. We consider electrons confined within a
hard-wall channel, of length $L$ and width $w$. The
$T_{ij}^{\alpha\beta}$ are calculated by injecting and following a
large number of electron trajectories (typically $>10^4$) propagating
at $v_F$~\cite{BB}.

For each path segment traversed by the electron between boundary
reflections, the phase of its spin wavefunction evolves continously
via the local Larmor frequency $\omega_L=g^*eB/2m$.  Here $g^*$ is the
effective electron $g$-factor, $B$ the local magnetic field, and $m$
the free electron mass.  Total precession is accumulated for each
complete trajectory, which is the sum of these segments.  For each
segment the electron's spin precession is calculated
analytically~\cite{17} and incorporated into the Monte Carlo
procedure.

In Fig.~\ref{fig2-3}(a) we display $R_S^{(\uparrow\uparrow)}$ as a function
of perpendicular magnetic field strength. The prominent and striking
new feature is that $R_S$ is {\it oscillatory,} a ballistic phenomenon
not found in the diffusive regime. In Fig.~\ref{fig2-3}(b,c) we display
$R_S^{(\uparrow\uparrow)}$ calculated for three orientations of the
external field---two that are in-plane and the perpendicular case,
displayed again for comparison. In all three cases the {\bf F1}, {\bf
F2} magnetizations are parallel and $\hat y$-oriented.

When the external field is along $\hat y$, the injected carriers
remain in spin eigenstates and do not precess.  In this situation
$R_S^{(\uparrow\uparrow)}$ is a positive constant [Fig.~\ref{fig2-3}(b,c)].
However, with an $\hat x$-oriented field precession is maximal, and
$R_S$ oscillates. Since orbital effects are absent for an in-plane
field, the oscillations in this case arise purely from spin precession
and the oscillation {\it period,} $\Delta B$, is determined by the
condition $2\pi n = \omega_L t_{TR}$, {\it i.e.}, $\Delta B =
h/(g^*\mu_B t_{TR})$. Here $t_{TR}=S/v_F$ is a typical transit time
from $1 \to 2$, and $\mu_B$ is the electronic Bohr magneton. $\Delta
B$ is thus inversely proportional to $S$, a typical path length
averaged over the injection distribution function. The {\it decay\/}
of $R_S$ occurs on a field scale where $\omega_L\delta t_{TR}\sim\pi$;
{\it i.e.}\ for $B=\hbar\pi/(g^*\mu_B\delta t_{TR})$ beyond which
precession amongst the different contributing trajectories tends to
get out of step. Here $\delta t_{TR} = [t_{TR}^2 -
\left<t_{TR}\right>^2]^{1/2}$ is the variance in path lengths
traversed while propagating from $1\to2$.

Perpendicular field $( B_{ext}||\hat z )$ is special---it induces both
spin and {\it orbital\/} effects. (The characteristic field scale for
the latter is $B_0=p_F/ew$, at which the cyclotron radius, $r_c = 
v_F/\omega_c$, equals the channel width, $w$.) The frequency ratio, 
$\omega_L/\omega_c = (g^*/2)(m^*/m)$ describes the relative importance 
of orbital and spin transport phenomena. Here, $p_F$ is the Fermi 
momentum, $\omega_c=eB/m^*$ the cyclotron frequency, and $m^*$ the 
effective mass.  For InAs ($m^*=0.025$, $g^*=15$) this ratio is $\sim0.19$, 
for InGaAs $\sim0.1$, whereas it is roughly 1.0 for most metals. In the 
latter spin and orbital effects have similar periodicity so disentangling 
them is difficult [Fig.~\ref{fig2-3}(a)].

\begin{figure}[t]
\begin{center}
\epsfig{file=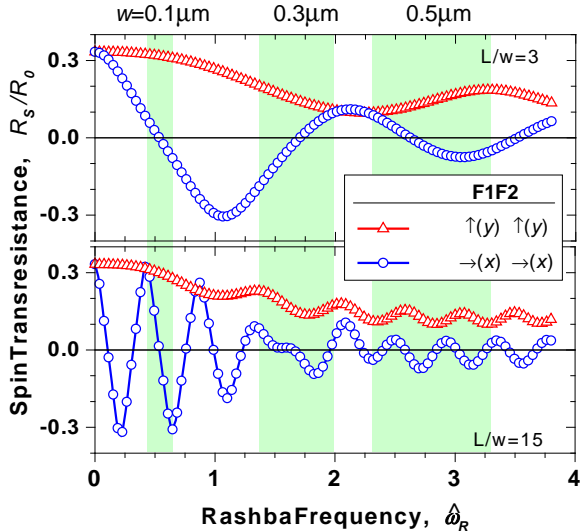,width=3.0in}
\end{center}
\caption{Spin transresistance {\it vs.} reduced Rashba frequency,
$\hat\omega_R= 2 m^* \alpha_{\rm so} w/\hbar^2$, at zero applied (external)
field, for two different device channel lengths, $L/w = 3$ and 15. The
parameter $\hat\omega_R$ can be controlled by an external gate
voltage. Shaded regions delineate the range of tunability expected for
InGaAs devices~\protect\cite{5} of three widths, 0.1, 0.3, and 0.5 $\mu$m.} 
\label{fig4}
\end{figure}

As mentioned, electrons confined within an InAs heterostructure are
subject to an internal Rashba field, present even for zero applied
magnetic field. This can be modeled by a Hamiltonian~\cite{4},
$H_R=\alpha_{\rm so}[\bbox{\sigma}\times{\bf k}]\cdot\hat z$.
Comparing $H_R$ to the Zeeman term we write the effective Rashba field
as ${\bf B}_R = {2\alpha_{\rm so}{\bf k}\times\hat z}/({g^* \mu_B})$.
Here $\alpha_{\rm so}=\Delta_R/2k_F$ is the spin-orbit coupling
parameter ($\Delta_R$ is the Rashba splitting~\cite{16}), and ${\bf
k}$ and $k_F$ are the electron and Fermi wave vectors,
respectively. Using data from Heida et al.~\cite{16}, we estimate this
internal field to be about 5T for an InAs 2DEG.  Since ${\bf B}_R$ is
always in-plane, all electron trajectories are straight when the
external field has no out-of-plane component.  For fully polarized
injection this yields the simple expression
\begin{equation}\label{eq5}
R_S^{(\uparrow\uparrow)} = 
{{1-2t}\over {(1+ 2t)(2t-3)}}
R_{sh},
\end{equation}
where $t$ represents $T_{(1\uparrow \to 2\uparrow)}$, normalized 
by $N_{ch}$. 

Figure~\ref{fig4} displays how Rashba-induced spin precession is
manifested in $R_S$ for zero external field. We represent the
effective Rashba field strength by the dimensionless frequency
$\hat\omega_R=2\alpha_{\rm so} m^* w/\hbar^2$; ~at $\hat\omega_R=1$ an
electron precesses one radian after traversing a distance $w$. As
shown, the oscillations decay quickly initially, but exceedingly
slowly thereafter.  No spin precession occurs for $\hat\omega_R=0$,
hence, $t=1$ yielding $R_S^{(\uparrow\uparrow)} = R_S^{(\to\to)} =
R_{sh}/3$, a simple result of current division. For finite
$\hat\omega_R$, $R_S$ displays strong dependence upon the orientation
of the magnetizations {\bf M} (of {\bf F1}, {\bf F2}; assumed
parallel), in relation to the device channel's principal axis ($\hat
x$). For ${\bf M}||\hat x$ (parallel to the channel), precessional
effects are maximal. With increasing Rashba field, the variance in
contributing path lengths causes the oscillations in $R_S$ to decay,
as described previously for the case of finite external field. Here,
however, the contributions from short paths (direct propagation
between the DSPR's involving few or no boundary reflections) continue
to add coherently for large $\hat\omega_R$, resulting very slow
decay. For ${\bf M}||\hat y$ most of the injected carriers experience
a Rashba field nearly aligned with their spin. At intermediate Rashba
field these yield small oscillations that center about a {\it
finite\/} value of $R_S$.  The other carriers make a contribution to
$R_S$ at small $\hat\omega_R$ but this becomes incoherent and thus
quickly decays for large $\hat\omega_R$~\cite{13,CC}.

The original idea of the spin transistor involved use of an external
gate potential, acting in concert with the intrinsic confinement
potential, to control of the spin precession rate~\cite{1}. We note,
however, that gate tuning of $\alpha_{\rm so}$ for electrons has been
experimentally demonstrated in relatively few narrow gap semiconductor
heterostructures. Two such systems are InGaAs/InP and
InGaAs/InAlAs~\cite{5,6}. In the latter, tuning over about a 30\%
range has been reported. In Fig.~\ref{fig4} we show how this range of
tunability translates into a direct modulation of $R_S$, for three
device widths. Our calculations clearly illustrate that the
``conventional'' spin transistor configuration, ${\bf M}||\hat y$,
(which is most easily fabricated) is {\it not\/} optimal---{\it even
for a very short channel\/} ($L\sim\ell_S$).  We find that tunability
is maximized for ${\bf M}||\hat x$.

The spin transistor was originally envisaged as a one-dimensional
device, with only a single populated transverse subband.  Realizable
devices in the near term will more likely be two-dimensional or,
perhaps, quasi-one dimensional channels.  Their increased phase space
for scattering can lead to quick suppression of $R_S$, especially in
the presence of moderate scattering~\cite{13}. Hence it appears that
an extremely narrow channel is a basic requirement for a spin
transistor.

Our calculations of $R_S$ clarify the important, and unique,
signatures of spin injected transport in an electron gas within a
semiconductor. They also point out crucial experimental challenges
that must be faced in making a spin transistor.

We gratefully acknowledge support from DARPA Spintronics, through ONR
grant number N00014-96-1-0865. We also thank A. Polichtchouk and
M.~Hartl for contributions to this work.


\begin{references}
\bibitem{7} An electro-optic modulator functions via crossed
polarizers that sandwich an electrically-tunable birefringent
material.
\bibitem{BB} We assume that the $T_{ij}^{\alpha\beta}$ are essentially
energy-independent on the scale of the spin splitting.  The latter is 
small compared to the Fermi energy in narrow gap, low density 
semiconductor heterostructures.
\bibitem{CC} We find that introduction of diffuse boundary or bulk
scattering to our calculations tends to wash out both the asymptotic
effects and the high field oscillations~\cite{13}. We also find that
our results are insensitive to the precise formulation of the DSPR's;
more complex (8-reservoir) calculations that include junction
scattering in these regions display similar physics~\cite{13}.
\end{references}
\end{document}